\documentstyle[aps,prl,epsfig,twocolumn]{revtex}
\begin{document}
%\twocolumn[\hsize\textwidth\columnwidth\hsize\csname @twocolumnfalse\endcsname
\preprint{IMSc 97/}
%\preprint{hep-th@ftp/9710020}
\title{Phase of the quantum oscillator}
\author{H.S. Sharatchandra \cite{email}} 
\address{The Institute of Mathematical Sciences, Chennai - 600 113, INDIA}
\maketitle
\begin{abstract}
Requirements of a conjugate operator are emphasized, especially
in its role in uncertainty relations.It is argued that in many
contexts it is necessary to extend the Hilbert space
in order to define a conjugate operator as in gauge theories.
Example of a particle in a box is analysed. This is closely 
related to the quantum oscillator through cosine states of 
Susskind and Glogower.It is used to justify London's phase wave
functions albeit as part of a larger Hilbert space.
A new definition phase uncertainty neccessiated by periodicity is
proposed.It is close to the usual r.m.s. definition.Corresponding 
number- phase uncertainty relation is 
obtained and its implications are discussed. Hilbert space of 
an oscillator is identified with the Hilbert space of a planar 
rotor with a $Z_2$ gauge invariance.This is
used to construct states analogous to the cosine and sine
states and to illustrate unitary equivalence of Hilbert
spaces.
\end{abstract}

\section{INTRODUCTION}
\label{intro}
There is a long history \cite{cn,m,rev,lyn} for attempts to 
define phase
operator for quantum oscillator and related systems.This is
important for quantum optics \cite{rev} at both theoretical and
experimental levels.We take a new look at this problem here.We 
argue that in many contexts it is necessary to extend the Hilbert
space in order to accomodate the conjugate operator as is done in
gauge theories. We use the close relation  between a particle in a
box and an oscillator provided by the cosine states of Susskind
and Glogower \cite{sg} to justify London's phase wave functions 
albeit
as part of a larger Hilbert space. This provides a very simple 
description of the oscillator states. Our strategy allows us
to obtain the number-phase uncertainty relation.We introduce a
new measure of uncertainty necessitated by periodicity. 
There are points of contact with many earlier works, especially
Refs. \cite{sg,n,gw,ll,g,st}.However our 
approach and results are different.

The paper is organized as follows. In sec. \ref{rotor} 
we  define a new measure of uncertainty of angular position
and obtain uncertainty relation in case of the planar rotor. This 
is directly related to the corresponding issues 
for a quantum oscillator.
In sec. \ref{box}, we point out the close relation
between a particle in a box and an oscillator provided 
by the cosine states of Susskind and Glogower \cite{sg}.
This is very useful to understand the conceptual issues 
and to resolve them. Therefore we analyse a particle in a
box as regards the conjugate momentum operator and 
uncertainty principle in sec. \ref{conc}. This shows 
that when the spectrum of an operator is bounded it is
necessary to extend the Hilbert space to accomodate the
conjugate operator and to obtain the expected consequences of 
Heisenberg uncertainty. We use these results to set the
requirements of the conjugate operator in sec. \ref{conj}.
The crucial requirement is that the inner product of the
eigenstates of the conjugate pair has to be a pure 
phase upto a constant normalization.Sometimes It is 
necessary to enlarge the Hilbert space to realize this.
This is a standard procedure used in gauge theories as
emphasised in sec. \ref{gauge}. We finally obtain the
phase wave functions of an oscillator and point out the 
simplicity they provide in sec. \ref{phase}.We discuss 
the number - phase uncertainty relation and its
implications in different contexts in sec. \ref{uncer}. 
In sec.\ref{z2} we point 
out that the Hilbert space of an oscillator can be 
regarded as that of a planar rotor with a $Z_2$ gauge
invariance. We use this to construct states similar to
the sine and cosine states \cite{sg} and illustrate the 
unitary equivalence  of Hilbert spaces.In particular the
oscillator states can also be represented by a particle in 
a box with antinodes at the walls.In sec. \ref{disc} 
we summarize our results. In Appendix  \ref{a} we consider some 
of the identities used. In Appendix \ref{b} we discuss
the definition of phase $\sl operator$.

\section{The planar rotor}
\label{rotor}
We first consider definition of uncertainty in angular position 
and the uncertainty relation for 
a quantum planar rotor. These are directly relevant for
the oscillator as discussed in later sections.

Angular momentum basis for a planar rotor is 
$|n>, n=0,\pm 1,\pm 2,\pm 3,\ldots$.
There is no problem in defining the phase eigenstates here.
They are,
\begin{eqnarray}
|\theta>&=&\sum_{-\infty}^{\infty} e^{-in\theta}|n>
\label{1}
\end{eqnarray}
\noindent
and represents the state in which the angular position on the
circle is specified with certainty.$|\theta>$ and
$|\theta+2N\pi>$ represent the same state so that the range
of $\theta$ may be restricted to $[-\pi, \pi)$.Planar 
angular momentum basis is normalizable, $<m|n>=\delta_{mn}$, 
whareas the phase basis $|\theta>$ is not.$|\theta>$ is not even 
a  well defined state of the Hilbert space.For example, the 
expansion coefficients in Eqn.  \ref{1} are not falling off with 
$n$. In a sense, $|\theta>$ 's are limit points of vectors of
the Hilbert space.Dirac has taught us how to handle this
profitably.  Orthonormality condition is now replaced by,
$<\theta|\phi>=\delta^P(\theta-\phi)$ where the periodic
Dirac delta function is
\begin{eqnarray}
\delta^P(\theta-\phi)
&=& \sum_{-\infty}^{+\infty}exp(-in(\theta-\phi)) 
\label{2}
\end{eqnarray}
An abstract state $|\Psi>$ may be completely specified through 
the wave function $\Psi(\theta)= <\theta|\Psi>$.
Allowed wave functions are given by normalizable and 
periodic functions of $\theta$. The inner product is,
\begin{eqnarray}
<\Psi|\Phi>&=&\int_{-\pi}^{\pi}\frac{d\theta}{2\pi}
\Psi^*(\theta) \ \Phi(\theta)
\label{3}
\end{eqnarray}
\noindent
Planar angular momentum operator $\hat L$ has the representation
$-i \hbar d/d\theta$ when acting on these wave functions as we
would expect of a conjugate operator.Thus the phase
eigenstates can be consistently defined in this problem
even though the conjugate variable has a discrete spectrum. 

\subsection{A new measure of phase uncertainty}
\label{meas}
If we choose the period $(-\pi+\alpha, \pi+\alpha)$, the
naive definition of uncertainty in phase is 
\begin{eqnarray}
\Delta_{\alpha}^2 \theta=\int_{-\pi+\alpha}^{\pi+\alpha} 
\frac{d \theta}{2 \pi}(\theta -< \theta>)^2 |\Psi(\theta)|^2
\label{4}
\end{eqnarray}
\noindent
where
\begin{eqnarray}
<\theta>=\int_{-\pi+\alpha}^{\pi+\alpha} 
\frac{d \theta}{2 \pi} \  \theta \  |\Psi(\theta)|^2
\label{5}
\end{eqnarray}
\noindent

There are serious problems with this choice, a consequence of
the periodicity of the wave functions.It depends on the choice
of $\alpha$, the origin chosen on the circle. Consider a narrow
wave packet centered around $\theta=\theta_0$ on the circle
and symmetric about it.
\begin{figure}[htb]
\begin{center}
\makebox[3.175in]{\epsfig{file=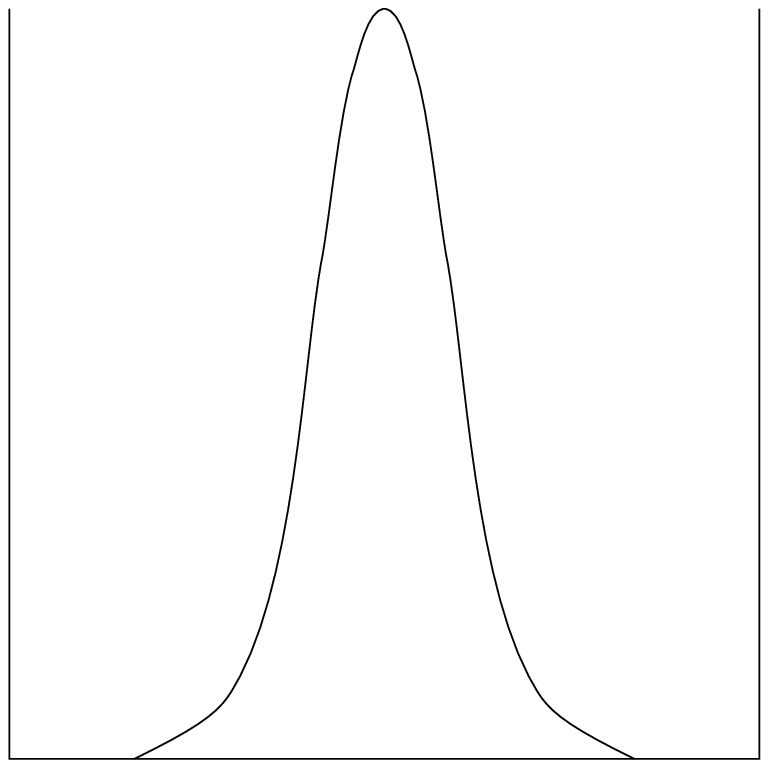,width=3cm,height=2cm,angle=0}
\hspace{1cm}\epsfig{file=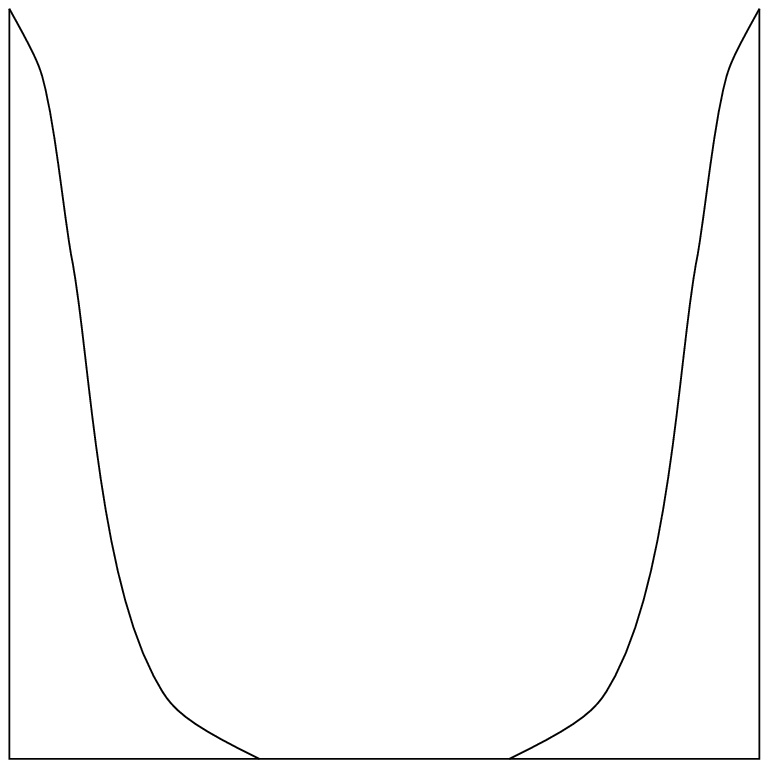,width=3cm,height=2cm,angle=0}}
\caption{a. Range chosen such that the wave packet is at the 
center b.Range chosen such that the wave packet is at the edge} 
\label{Fig 1}
\end{center}
\end{figure}
If we choose $\alpha=\theta_0$, (Fig.1a), we get
$<\theta>=0$ as a consequence of symmetry. Then $\theta-<\theta>$
is small at the wave packet and we get a small uncertainty
in angular position as one would expect for a narrow wave packet.
On the other hand if we choose $\alpha=\theta_0+\pi$, (Fig.1b),
again $<\theta>=0$ due to symmetry. Now $(\theta-\theta_0)^2
\approx \pi^2$ at the wave packet and we get a large uncertainty
which is not acceptable.We need a definition which automatically 
makes the earlier choice.

The usual definition of uncertainty of a variable can be 
reinterpreted as follows. Consider
\begin{eqnarray}
\int d \theta \  (\theta-\theta_1)^2  |\Psi(\theta)|^2
\label{6}
\end{eqnarray}
\noindent
which is measuring the mean of square of deviation measured from 
an arbitrary  point $\theta_1$. This is minimum when
$\theta_1=<\theta>$. Thus we are minimizing the mean square 
of the distance measured from an arbitrary point.

We adopt this principle for our case. Thus our choice is
\begin{eqnarray}
\Delta^2 \theta=min \int_{-\pi+\alpha}^{\pi+\alpha}\frac{d\theta}
{2\pi} \ (\theta -<\theta>)^2 \ |\Psi(\theta)|^2
\label{7}
\end{eqnarray}
\noindent
where we now minimize w.r.t the parameter $\alpha$ which could be
over the entire range $(-\infty, \infty)$.  Extremization gives
\begin{eqnarray}
0 &=& 2(\alpha-< \theta>) |\Psi(\pi+\alpha)|^2 \nonumber \\
&&-\int_{-\pi+\alpha}^{\pi+\alpha}\frac{d\theta}
{2\pi} 2(\theta -<\theta>) \frac{d<\theta>}{d\alpha}\ 
|\Psi(\theta)|^2
\label{8}
\end{eqnarray}
\noindent
noting that $<\theta>$, Eqn. \ref{5}, itself is a function of $\alpha$.
We have $d<\theta>/d\alpha=|\Psi(\pi+\alpha)|^2$ so that the second
term is zero. Therefore Eqn. \ref{8}
gives the extremization condition to be $<\theta>=\alpha$.
This simply means that we must choose the range such that
the mean value of $\theta$ is at the center of that range.
There can be many local extrema and therefore many choices of 
the range satisfying this condition.  Some of them
could correspond to local maxima also. The second derivative 
w.r.t $\alpha$ is
\begin{eqnarray}
2(1-\frac{d<\theta>}{d\alpha})|\Psi(\pi+\alpha)|^2+
2(\alpha-<\theta>) \frac{d|\Psi(\pi+\alpha)|^2}{d\alpha}
\label{9}
\end{eqnarray}
\noindent
The second term is zero at the extrema.
Thus the condition for minima  is $|\Psi(\pi+\alpha)|^2<1$ i.e.
the wave function at the edge of the interval should have 
a magnitude less than one.This is realized in Fig. 1a. On the
other hand if the magnitude is greater than one as in Fig. 1b we
get a local maxima.In principle there could be more than one
minimum and we must choose the absolute minimum.

It is illustrative to present formula for the phase uncertainty
in the momentum basis.If $\Psi(\theta)=\sum c_l \ exp(il\theta)$
with normalization condition $ \sum |c_l|^2 = 1$, then
\begin{eqnarray}
<\theta>&=&\alpha+\sum_{l,L \neq l} c_l^* \  c_L
\frac{(-1)^{l-L}}{-i(l-L)} e^{-i(l-L) \alpha}  \nonumber \\
<\theta^2>&=&\frac{\pi^2}{3}-\alpha^2+ 2\alpha <\theta> \nonumber\\
& & +2\sum_{l,L \neq l} c_l^* \ c_L
\frac{(-1)^{l-L}}{(l-L)^2} e^{-i(l-L) \alpha} 
\label{9a}
\end{eqnarray}
\noindent
Thus the extremization condition $<\theta>=\alpha$ is
\begin{eqnarray}
0=\sum_{l,L \neq l} |c_l \  c_L|
\frac{(-1)^{l-L}}{(l-L)} sin((l-L) \alpha-\beta_l+\beta_L)
\label{9b}
\end{eqnarray}
\noindent
where $c_l=|c_l|exp(-i\beta_l)$.At these extrema,
\begin{eqnarray}
\Delta_{\alpha}^2 \theta=\frac{\pi^2}{3}
+\sum_{l,L \neq l} |c_l \  c_L|
\frac{2(-1)^{l-L}}{(l-L)^2} cos((l-L) \alpha
-\beta_l+\beta_L)  \nonumber \\
\label{9c}
\end{eqnarray}
\noindent
The condition for minimization is
$\sum_{l,L \neq l} (-1)^{l-L} |c_l c_L| cos((l-L)\alpha
-\beta_l+\beta_L) \leq 0$. Note that r.h.s of Eqn. \ref{9c}
is a periodic function of $\alpha$, and minimization condition
simply corresponds to the minimum of this function.Further 
this minimum is necessarily less than $\pi^2/3$.This is because 
the cosines all have non-zero Fourier modes so the the mean
value of $\Delta_{\alpha}^2 \theta$ is $\pi^2/3$.

Our definition of the phase uncertainty satisfies all intuitive
requirements. It also leads to a simple uncertainty
principle of the Heisenberg type as seen below.

\subsection{Uncertainty relation}
\label{un}
In an angular momentum eigenstate, the angular position
probability is uniformly spread over the entire circle.
This is as expected of the conjugate operator.However
the way this comes about is somewhat novel.

The uncertainty relation can be worked out in the
in the usual way by starting with the inequality
$||(\bar L+ir \bar \theta) \Psi|| \geq 0$ where $r$ is
a parameter to be varied to get the best inequality. 
Here $\bar X \equiv \hat X-<\hat X>$ for any operator
$\hat X$. To begin with we will choose an arbitrary range.
Note that, in general, $\theta \Psi(\theta)$ has unequal values 
at the boundary of the interval even though $\Psi(\theta)$ is 
periodic.As a consequence the boundary contribution
\begin{eqnarray}
\int_{-\pi+\alpha}^{\pi+\alpha}\frac{d\theta}{2\pi}
\ \frac{d}{d\theta}(\theta |\Psi(\theta)|^2)
\label{10}
\end{eqnarray}
\noindent
is not zero but $|\Psi(\pi+\alpha)|^2$.This gives a new 
contribution to the uncertainty relation:
\begin{eqnarray}
\Delta L \  \Delta_{\alpha} \theta \geq
\frac {\hbar}{2} |(1-| \Psi(\pi+\alpha)|^2)|
\label{11}
\end{eqnarray}
\noindent
This relation can be derived directly from operator techniques
(\ref{b}).Note that if the range chosen is such that the
wave function $\Psi(\pi+\alpha)$ at the edge has magnitude
larger than one then the r.h.s. is $|\Psi(\pi+\alpha)|^2-1$.
This can be arbitrarily large.Also it
can become arbitraily close to zero.

Finally by considering this at the value $\alpha = \alpha_0$ which
minimizes  $\Delta_{\alpha} \theta$ we get the uncertainty 
relation between fluctuations in phase and planar angular
momentum:
\begin{eqnarray}
\Delta L \ \Delta \theta \geq
\frac {\hbar}{2} (1-| \Psi(\pi+\alpha_0)|^2)
\label{12}
\end{eqnarray}
\noindent
Note that
in this case we are assured that $|\Psi(\pi+\alpha_0)| \leq 1$.

Now we consider various implications of this equation.
i. As the spectrum of $\hat L$ is discrete, we can now have
a normalizable state $exp(il\theta)$ with $\Delta L=0$.
In absence of the extra term on the r.h.s.,Eqn. \ref{11}, 
$\Delta \theta$ would then be unbounded. This is not possible 
because the  range  of $\theta$ is finite.
This contradiction is avoided because of the novel form
of the r.h.s. For the angular momentum eigenstate, 
$\Delta L=0$ and $|\Psi(\theta)|=1$, so that both
sides of Eqn. \ref{11} are zero.The uncertainty in phase is
$\pi/\sqrt 3$ as expected for a random phase distribution
\cite{bp1}.Indeed this is the maximum value of phase uncertainty
with our definition.We may see this as follows.
We may expect large phase uncertainty for two identical wave 
packets that are diametrically opposite on the circle. In this 
case $\theta$ has to be measured from the midpoint of the 
two on the circle.Of the two midpoints,  we have to choose the
one for which the distance is shorter.Thus the phase uncertainty
is made larger by choosing the wave packets to be symmetric 
about the diametrically opposite points.Consider such 
normalized wave packets of width $\delta$:
$|\Psi( \theta)|^2=\pi/\delta$ for $\theta \in (\pi/2-\delta/2 ,
\pi/2+\delta/2) and  (-\pi/2-\delta/2, -\pi/2+\delta/2)$.
The maximum value of $\delta$ is $\pi$ which corresponds to
a uniform probability over the entire circle.
We have $\Delta ^2 \theta=\pi^2/4+\delta^2/12$. 
This is maximized for $\delta=\pi$, i.e. for a uniform 
probability and the maximum value is $\pi^2/3$.

ii. We now consider how the inequality is satisfied for
a narrow normalized wave packet 
\begin{eqnarray}
\Psi( \theta)=\sqrt{tanh \ \epsilon}
\ \sum_{-\infty}^{\infty}e^{-|l|\epsilon+il(\theta-\beta)}
\label{13}
\end{eqnarray}
\noindent
which becomes the periodic delta function at $\theta=\beta$ in 
the limit $\epsilon \rightarrow 0$ (Appendix \ref{a}).Now 
\begin{eqnarray}
\Psi(\theta)=\sqrt {\frac{(1-e^{-2\epsilon})^3}
{(1+e^{-2\epsilon})}}
\frac{1}{(1-e^{-\epsilon})^2+4 sin^2((\theta-\beta)/2)}
\label{14}
\end{eqnarray}
\noindent
This is a wave packet of scale $O(\epsilon)$.Since it is peaked 
symmetrically about $\theta=\beta$, we must choose the range
$(-\pi+\beta, \pi+\beta)$ for calculating uncertainty in $\theta$
according to our prescription, Eqn \ref{7}.
For small $\epsilon$ we get by a rescaling $\theta=x\epsilon$,
\begin{eqnarray}
\Delta^2 \theta = 4\epsilon^2 \int_{-\infty}^{\infty} 
\frac{dx}{2\pi} \frac{x^2}{(1+x^2)^2}+O(\epsilon^3)
\label{15}
\end{eqnarray}
\noindent
so that $\Delta \theta =\epsilon+O(\epsilon^2)$ as one would 
expect for a wave packet 
of width $\epsilon$.Thus the phase uncertainty can be made
arbitrarily small.Since $<L>=0$, we
have $\Delta L = <L^2>$ and 
$\Delta L = \sqrt 2 \hbar (e^{\epsilon}-e^{-\epsilon})^{-1}$.
Therefore $\Delta L = \hbar/(\sqrt 2  \ \epsilon +O(\epsilon))$
diverges as $1/ \epsilon$ as for a wave packet of size $\epsilon$
in position space.We have  $\Delta L \ \Delta \theta =
\hbar/\sqrt 2+O(\epsilon)$ . This satisfies the 
uncertainty relation Eqn. \ref{12} because the wave function
at the edge of the interval $\Psi(\beta+\pi) =O( \sqrt 
{\epsilon^3})$ is very small.

iii. We now consider a normalized state
\begin{eqnarray}
\Psi( \theta) &=&
cos \gamma \ e^{il\theta}+sin \gamma \ e^{-i\beta}e^{iL\theta}
\label{17}
\end{eqnarray}
\noindent
for $l \neq L$.
For small $\gamma$ this is dominated by the angular momentum 
state $l$ with a small admixture of angular momentum $L$.
The extremization condition gives $sin((l-L)\alpha+\beta) = 0$.
We get $\Delta^2 \theta=\pi^2/3-2(l-L)^{-2} |sin(2\alpha)|$.
On the other hand, $\Delta L = |(l-L) sin(2\alpha)| \hbar/2$. 
Thus, $\Delta L \  \Delta \theta = | sin(2\alpha)| 
\sqrt{\pi^2 (l-L)^2/3-2|sin(2\alpha)|} \hbar/2$.This is 
always larger than the r.h.s. of our uncertainty relation which 
is  now $|sin(2\alpha)| \hbar/2$.

\section{ Analogy to a particle in a box}
\label{box}
In this section we point out that the cosine states of
Susskind and Glogower \cite{sg} relate the stationary 
states of 
an oscillator and a particle in a box.We use this for
understanding some conceptual issues regarding the phase
operator of an oscillator in the next section.

Eigenstates of the number operator $\hat N$ of an
oscillator are denoted by $|n>, n=0,1,2, \ldots$.
The label is truncated from below.This is the cause
underlying the difficulties in defining the conjugate
phase operator.  Susskind and Glogower \cite{sg} 
(henceforth to be referred to as SG) constructed 
eigenstates of the `cosine operator' 
$\hat c=\sum_1^{\infty} (|n><n-1|+ |n-1><n|)$.
This operator is self adjoint and has a continuum set of 
eigenstates (in the generalized sense). We  label them
$|\theta \supset$.(We are using a different notation than
the usual ket only to distinguish this state from the
angular position eigenstate $|\theta>$ defined in the
sec.\ref{rotor}.) The corresponding eigen value is 
$ cos \theta $.
\begin{eqnarray}
|\theta\supset  &=& \sum_{0}^{\infty} sin((n+1) \theta) |n>
\label{20}
\end{eqnarray}
\noindent
These states are not well defined states of the Fock space
as with the states in Eqn. \ref{1}. They have to be
handled in the same way as using the Dirac formalism.
Being eigenstates of a Hermitian operator they form a 
complete orthonormal set.
The inner product of two such states is, (see Appendix \ref{a})
\begin{eqnarray}
\subset \theta|\phi \supset  &=& \sum_{0}^{\infty} 2sin((n+1) 
\phi) sin((n+1) \theta) \nonumber \\ 
&=& \delta^P(\phi-\theta) -\delta^P(\phi+\theta)
\label{21}
\end{eqnarray}
\noindent
$|-\theta>$ is the same as $-|\theta>$ and it suffices to 
choose the range $[0,\pi)$ for $\theta$.
Then the second $\delta$ function drops out and we get the
Dirac delta function orthonormality.

Instead of $<\theta|\Psi>$ we may use the wave function 
$\subset \theta| \Psi>$  to describe the states of an 
oscillator.Then the  stationary state $|n>$ of an oscillator has 
the wave function $sin((n+1) \theta)$.
Let us presume that there is an apparatus to measure the
cosine operator.Then the stationary states of an
oscillator are exactly those of a particle in a box in
the usual Schroedinger description.Thus a reformulation
of an oscillator using the cosine operator in place of the
position operator maps its eigenstates  to that of a particle 
in a box.  (This is discussed further in sec. \ref{z2}).

\section{Some conceptual issues}
\label{conc}
In this section we analyse a particle in a box to resolve 
some conceptual issues regarding the conjugate operator
and uncertainty relation.

Consider a particle in a box with the position $x$
in the range $[0, \pi)$. The states of the system are
linear combinations of $sin((n+1)x), n= 0,1,2,\ldots$.
It appears that the momentum operator $-i \hbar
d/dx$ is not defined on the Hilbert space. It would
take a sine function to a cosine function which is not a
state of the Hilbert space. This raises the question as
to  whether the Heisenberg uncertainty relation can be
applied to the system.The text book derivation of the 
zero point energy of a particle in box uses precisely
the Heisenberg uncertainty. A formal way of justifying
this is as follows. Enlarge the Hilbert space to one
spanned by the plane waves $exp(\pm i nx), n=1,2,3, \ldots$
The momentum operator $-i \hbar d/dx$ is well defined on
this enlarged Hilbert space.If we include $n=0$ wavefunction also 
we get precisely the planar rotor basis, Eqn. \ref{1}.We may 
therefore apply the uncertainty relation Eqn.\ref{12}.However we 
use it only for the `physical states' of the particle in a box
i.e. those spanned by the sine functions only.In particular this
means that the state with $\Delta L=0$ is not an allowed state.

To obtain uncertainty relation we regard the wave functions of 
the particle in a box as a subset of wave functions of the planar 
rotor. $x \subset (0,\pi)$ is identified with $\theta \subset
[-\pi,\pi)$.The wave functions are periodic repetitions of the
wave functions of a particle in a box.They vanish at $\theta
=\pi(mod 2\pi)$. The uncertainty in position is define as in 
sec. \ref{meas}. For all `physical
states' the probability for angular momenta $\pm L$ are
equal. Therefore $<\hat L>=0$ and $\Delta^2 L=<\hat L^2>$. Now,
$\hat L^2$ is a meaningful operator on the physical states, as it
takes the sine functions into sine functions.It is precisely  the
Hamiltonian of the system.Thus the uncertainty relation as
obtained by extending the Hilbert space constrains the expectation 
value of energy and the spread in position.In particular it gives
a zero point energy.

The lesson to be learnt from this example is that when the 
spectrum of an oscillator is bounded, it is necessary to extend
the Hilbert space and regard the states of the original Hilbert
space as `physical states' of the system. Only then it is 
possible to define the conjugate operator and obtain implications
of the uncertainty relations for the original operator.
 
\section{Requirements of a conjugate operator}
\label{conj}
Our discussion of the planar rotor and a particle in a box is
to emphasize that the definition of the conjugate operator is
more involved when the spectrum of the original operator is
discrete or compact. In this section we formally state the
properties required of a conjugate operator and the neccessity
to enlarge the Hilbert space at times to accomodate it.

In case of the planar rotor, angular momentum operator has the 
traditional form $-i \hbar d/d\theta$ when acting on the wave
functions $\Psi(\theta)$. But its commutator with the 
angular position operator (Appendix \ref{b}) is not the standard 
one.  Nevertheless expectations of the Heisenberg 
uncertainty principle are in operation. In the angular position
eigenstates $exp(il\theta)$,  the angular position probability 
is spread uniformly over all angles as the probability amplitude 
is just a phase. Consider a normalized wave packet peaked around
$\theta_0$ on the circle. In the limit of an infinitely narrow
wave packet, it is the periodic delta-function  $\delta^P(\theta-
\theta_0)$ upto a diverging normalization.The probability
amplitude to find angular momentum $l$ is proportional to
$exp(-il\theta_0)$, Eqn. \ref{2}, and hence there is a uniform 
spread in all
angular momenta. Thus in this example the crucial feature of the 
conjugate operator that is relevant for the
Heisenberg uncertainty principle is not the canonical Heisenberg 
commutation relation.  It is the fact that the matrix elements
$<\theta|l>$ of the eigenstates of conjugate operators are all
phase factors upto an overall normalization.

Now consider a particle in a box.
A wave packet which is localized around a point $x_0$
can be expanded in the  complete set, the sine
functions. In the limit of the wave packet becoming infinitely
narrow, we get a representation (upto an infinite normalization)
of the position eigenstate for the problem(Appendix \ref{a}).
$\Psi (x)\equiv \sum_0^{\infty} sin((n+1)x_0)
sin((n+1)x)$. The probability $ sin^2((n+1)x_0)$ for 
different $n$'s are not of same magnitude. But these coefficients 
refer to an expansion in the eigenstates of the Hamiltonian 
and not of conjugate momentum operator.To get the eigenstates 
of the latter we enlarged the Hilbert space in sec.\ref{conc}. 
Now in this basis the 
position eigenfunction has the expansion $\Psi \subset x
\supset = \sum_{-\infty}^{\infty} exp(inx_0) exp(-inx)$  as in 
the planar rotor.
Now the coefficients are all phase factors and the probability is
spread uniformly over all momenta. Thus again the crucial feature
of the conjugate operator is that the matrix element 
is a phase. We had to enlarge the Hilbert space to accomodate the
conjugate operator in this case.

Consider a self adjoint operator $\hat A$ with a complete 
orthonormal set of
eigenstates $|\alpha>$ with real eigenvalues $\alpha$. This may 
be a continuum set or discrete;  of a bounded or unbounded 
spectrum;
or a combination of these. The conjugate operator $\hat B$ is 
required to have the following properties in order to realize 
the expectations from the Heisenberg uncertainty principle.
i. It is a self adjoint operator with a complete orthonormal set
of eigenstates $|\beta>$ with real eigenvalues $\beta$. ii.The
matrix elements $<\beta|\alpha>$ should all be phase factors upto
a normalization independent of the labels $\alpha$ or $\beta$.
The original Hilbert space may not admit an operator with these
propertes. Then we have to enlarge the Hilbert space and regard
the states of the original Hilbert space as `physical states'. We
consider further examples of this in latter sections.

\section{Analogy with gauge theories}
\label{gauge}
In this section we point out that our technique of enlarging the
Hilbert space is the standard procedure used in gauge
theories.Therefore various techniques of quantization used in
gauge theories can be adopted for our systems.

In quantum electrodynamics, to get a Hamiltonian description we 
are forced to introduce an  unphysical  operator $A_i(x)$, 
the vector potential,  as the conjugate of $E_i(x)$,the electric 
field.  The gauge invariant operator, magnetic field $B_i(x)$,  
has more complicated commutation relations 
with $E_i(x)$.If we are using a Schroedinger 
wave functional description $\Psi[A_i(x)]$, not all wave 
functionals
are in the physical Hilbert space. Only such which have same
values for any given $A_i(x)$ and $A_i(x)+ \nabla_i \lambda(x)$ 
(where the function $\lambda(x)$ is arbitrary) are physical 
states.

As we are going to use the analogy with quantum electrodynamics, 
we give a quick review of the various quantization procedures used 
there.  i)We may work with only the
subspace of the Hilbert space consisting of physical states, those
that satisfy the Gauss law constraint, $\nabla_i E_i(x)|phys>=0$.
ii)We may choose a representative from gauge equivalent
configurations.A common choice is the Coulomb gauge  fixing 
condition $\nabla_i A_i=0$.Then arbitrary functionals $\Psi[A_i]$ 
of such $A_i(x)$ may be used.However the canonical commutation
relation may have to be modified to account for the gauge fixing
constraint.iii)We may eliminate the redundant degrees altogether,
as in the axial gauge where $A_3(x)$ is eliminated and only
$A_1,A_2$ are used.iv)We may use the electric field basis and
solve the Gauss law.v)There is also the BRST quantization 
involving additional Grassmann degrees of freedom.

Each of the above procedures may be adopted for the present
situation.They would all give same results though a particular
representation may be advantageous for a given purpose. We 
directly pass on 
to the the most convenient choices  for the present case.

\section{Phase wavefunctions of the quantum oscillator}
\label{phase}
In sec. \ref{conj} we have argued that the crucial 
requirement for the conjugate phase description of the number
operator eigenstates is that the wave function $< \theta|n>$
of the number eigenstate should be phase factors upto a 
normalization so that it has uniform probability for each 
value of the phase. Many authors,
in particular SG \cite{sg}, have pointed out 
problems with this.If we define the phase eigenstates,
\begin{eqnarray}
|\theta )) =\sum_0^\infty \; e^{-i n\theta} |n>
\label{22}
\end{eqnarray}
\noindent
with the sum running over only the non-negetive integers,
then the states are neither linearly independent nor orthonormal:
The inner product is (Appendix \ref{a}),
\begin{eqnarray}
(( \theta|\phi ))=\frac{1}{2}\delta^P(\theta-\phi) 
+i e^{i(\theta-\phi)/2} \;  cosec(\theta-\phi)/2
\label{23}
\end{eqnarray}
\noindent
The way out of this problem is suggested by our analogy of a 
particle in a box. We enlarge the 
Hilbert space though it is now used in a different way.
This Large Hilbert space has the basis $|n>$ where $n$
now takes both positive and negetive integral values.The
physical states are only those with $n= 0,1,2, \ldots$.
The purpose of having the additional states is to be able
to define the conjugate phase operator.The basis of the
Large Hilbert space is in 1:1 correspondence with the
angular momentum basis of the planar rotor and we may
define the conjugate operator $\hat \theta$, Eqn. \ref{38},
and its eigenstates $|\theta>$, Eqn. \ref{1}.These states
are free of the problems  encountered by SG
\cite{sg}. The
eigenstate $|\theta>$,  Eqn. \ref{1}, is built out of both
positive and negetive $n$ states and is therefore not a 
physical state.  But this does not mean that we cannot use 
it to describe the states of the subspace.

The analogy with the gauge theories is as follows.We may 
describe the physical states of quantum electrodynamics using 
the wave functional $<\{A_i(x)\}|\Psi>$ even though the gauge 
potential $\hat A_i(x)$ is not an operator on the physical
subspace. Only, not all wave functionals are allowed but just
those satisfying the Gauss constraint, 
$\nabla_i \delta \Psi(A)/\delta A_i(x)=0 $. The analogous 
constraint for us is that the physical states should be
composed of only non-negetive values of the angular momentum 
$-i \hbar d/d\theta$.This has some analogy to the Gupta-Bleuler
technique in quantum electrodynamics.

As in gauge theories, even though $|\theta>$ is not a 
physical state, states  of the physical Hilbert space 
may be described via the wave functions $\Psi(\theta)=
<\theta|\Psi>$ and all observables may be
computed from it.For a state with the occupation number 
expansion $ |\Psi>=\sum_0^{\infty} c_n |n> $ the 
corresponding phase wave function is $\Psi(\theta)=
\sum_0^{\infty} c_n \ exp(in \theta)$. This is precisely
the phase representation of state vectors implicit in 
papers of London \cite{l} and often used in quantum optics
\cite{qo1,qo2,qo3}.  The major advantage of our point
of viewing it as a subspace of the Large
Hilbert space is that we can obtain the number - phase
uncertainty relation and its implications. This is 
discussed in the next section. Note that it is not 
possible to resolve a general wave packet using 
positive Fourier modes only.(In particular, the $\delta$-
function cannot be resolved and we cannot realize the
phase eigenstate.) We construct other physical 
states of narrow phase distribution in the next section.

Working with the Large Hilbert space provides a 
simplification both conceptually and computationally.
The time evolution of a state has a simple form when
the phase description is used. $\Psi(\theta) 
\rightarrow \Psi(\theta- \omega t)$ is obtained by the
classical evolution $\theta \rightarrow \theta - \omega t$
as expected.Inner product of any two wave functions 
$\Psi(\theta)$ and $\Phi(\theta)$ is given by Eqn. \ref{3}. 
The annihilation operator has the simple representation
$\hat a=e^{-i\theta}\sqrt{-id/d\theta}$ when acting on the 
phase wave functions.  Note that by itself the multiplication by 
$e^{-i \theta}$ takes the ground state of the
oscillator outside the physical subspace and therefore not
a physical operator.But the operator $\sqrt{-id/d\theta}$ 
gives zero precisely on this state making $ \hat a$ a physical 
operator.Thus the Dirac's decomposition of the annihilation 
and creation operators is now valid. Using the Large Hilbert 
space has been crucial for making this possible.The contradiction 
noticed by Louiselle \cite{lou} and SG \cite{sg} 
is not valid now.  $\hat u=exp(-i\theta)$ satisfies 
$\hat u \hat u^*=\hat u^* \hat u =\bf{1}$ because now it is
an operator in the Large Hilbert space.

\section{Number-phase uncertainty relation}
\label{uncer}
We now consider the number-phase uncertainty relations
for the quantum oscillator.As the phase wave functions 
are a subset of the wave functions of a planar rotor
we may directly use the uncertainty relation Eqn. \ref{12}.
But the following points should be
kept in mind.We use the uncertainty relation for physical
states only, though it is valid for other states also.
$\hat \theta$ (Appendix \ref{b}) is not a `physical' operator. It 
generates negetive Fourier modes also when acting on a physical 
state.  Nevertheless the uncertainty relation
Eqn. \ref{12} gives a constraint on the relative spreads in 
the occupation number distribution ( represented by $|c_n|^2$)
$vis-a-vis$ the phase (represented by $|\Psi(\theta)|^2$).For 
this purpose the r.m.s. spread in $\theta$, Eqn. \ref{7}, is 
again relevant and meaningful as it stands. We may express 
this in stronger words as follows: The wave function
$\Psi(\theta)$ contains complete information of the
abstract state $|\Psi>$ and therefore meaningful even if 
$|\theta>$ is not a physical state. So long as the states 
are described by
the wave functions $\Psi (\theta)$, $\Delta \theta$
as defined in Eqn. \ref{7} is calculable and therefore
a measurable object. Further it a measure of the spread in
$\theta$. Therefore the number -phase
uncertainty relation is given by Eqn \ref{12} with
$\Delta L/ \hbar$ being interpreted as the r.m.s.
spread in the number.

The occupation number eigenstate $|n>$ has a uniform  
spread  in the phase as expected and meets the `acid test' 
of Barnett and Pegg \cite{bp1}.On the other hand the phase 
eigenstate $|\theta>$ is not a physical eigenstate and 
therefore the discussion concerning it is not relevant for 
the oscillator.

We now consider states \cite{m,ll,ss} which are 
sometimes referred to as coherent phase states:
$\Psi_{\zeta}(\theta)=\sqrt{1-|\zeta|^2}
\sum_0^{\infty}\zeta^n \ exp(in\theta)$.These are obtained as 
eigenstates of the non-Hermitian `exponential operator'
$E=\sum_0^{\infty}|n><n+1|$.This is formally similar to
the  narrow wave packets of the planar rotor Eqn. \ref{13} 
with $\zeta \rightarrow exp(-\epsilon-i\beta)$, but  only
positive Fourier modes are involved.
Therefore it is of interest to see whether they correspond to
states of the oscillator with a narrow phase distribution. 
Note that with $\zeta=exp(-\epsilon-i\beta)$,
\begin{eqnarray}
|\Psi_{\zeta}(\theta)|^2
&=& (1-e^{-2\epsilon})\sum_{m,n=0}^{\infty} 
e^{-(m+n)\epsilon} e^{i(n-m)(\theta-\beta)} \nonumber \\
&=& (1-e^{-2\epsilon}) \sum_{N=0}^{\infty}
e^{-2 N\epsilon} \sum_{r=-\infty}^{+\infty} e^{-r\epsilon}
e^{ir(\theta-\beta)} 
\label{24}
\end{eqnarray}
This, ofcourse, has both positive and negetive 
Fourier modes. Moreover it is precisely the wave packet 
of the planar rotor, Eqn. \ref{13}, considered as an 
approximation to the periodic delta function function
(Appendix \ref{a}). For $\epsilon  \rightarrow 0$,  
$\Psi_{\zeta}(\theta)$ is sharply peaked 
in $\theta$ around $\theta=\beta$. Thus it is a state
of narrow phase distribution.It is a complex square root
of the Poisson kernel (Appendix \ref{a}), and in the limit 
$\zeta \rightarrow 1$ it may be regarded as square root of the
periodic delta function. 

As the probability density is peaked and symmetric about
$\theta=\beta$, we expect that we have to choose $\alpha=\beta$,
Eqn. \ref{7}, for the uncertainty in phase. Therefore we get
\begin{eqnarray}
\Delta^2 (\theta)&=& (1-|\zeta|^2)
\int_{-\pi}^{\pi} \frac{d\theta}{2\pi}
\frac{\theta^2}{(1-|\zeta|)^2+4 |\zeta| sin^2(\theta/2)}
\label{25}
\end{eqnarray}
The integral is finite in the limit $|\zeta| \rightarrow 1$
and has the value  $ ln \  4$. Therefore $\Delta \theta
\rightarrow 0$, and we get a vanishingly 
small uncertainty in phase. On the other hand,
$<\hat N>=\hbar |\zeta|(1-|\zeta|^2)^{-1}, <\hat N^2>=\hbar^2 
|\zeta| (1+\zeta|)(1-|\zeta|)^{-3}$ so that
$\Delta N =  |\zeta|^2(1-|\zeta|^2)^{-1}$.
Note the following features.As $\epsilon \rightarrow 0$, 
$\Delta \theta =2 \sqrt {ln2} \sqrt \epsilon+O(\epsilon)$.Thus 
the uncertainty in $\theta$ is much larger $O(\sqrt
\epsilon)$ than the scale $O(\epsilon)$ of the wave
packet, in contrast to the free particle or the wave
packet, Eqn. \ref{13}, for the planar rotor. The reason is
that $\Psi_{\zeta}(\theta)$ is quite wide.It is however true
that we can obtain states of arbitrarily small
uncertainty in phase by choosing $\epsilon$ arbitrarily
small. In this state $\Delta N \sim (2 \epsilon)^{-1}$ diverges 
as is to be expected. 
The product $\Delta N \Delta \theta \sim 1/\sqrt \epsilon$ 
is much larger than the r.h.s. of the uncertainty relation. Note 
that $|\Psi(\pi+\beta)|^2$ is vanishingly small in this limit
so that the extra term on the r.h.s. of the uncertainty
relation Eqn. \ref{13} has almost no effect.

We now consider the phase uncertainty for coherent state $|z>$ 
which has the normalized phase wave function
\begin{eqnarray}
\Psi_z (\theta ) &=& e^{-r^2/2} \sum_{0}^{\infty} 
\frac{r^n}{\sqrt {n!}} e^{in(\theta-\beta)}
\label{27}
\end{eqnarray}
\noindent
where $z=r \ exp(-i\beta)$.This is symmetrically peaked 
around $\theta=\beta$.Therefore we have to choose the
range $(-\pi+\beta, \pi+\beta)$.(This can be checked directly.) 
The uncertainty in phase $\Delta \theta$ is
\begin{eqnarray}
\Delta^2 \theta &=& \frac{\pi^2}{3}+
e^{-r^2} \sum_{m,n \neq m} \frac{2(-r)^{m+n}}
{\sqrt {m! n!}(m-n)^2} \nonumber \\
&=& 4e^{-r^2} \sum_{n=0}^{\infty} \frac{r^{2n}}{\sqrt{n!}}
\sum_{s=1}^{\infty} \frac{(-r)^s}{s^2}(\frac{1}{\sqrt{(n+s)!}}
-\frac{1} {\sqrt{n!}})
\label{28}
\end{eqnarray}
\noindent
This is small for large $r$.This can be directly seen from the 
form of $|\Psi(\theta)|^2$. The half width is $O(r^{-1})$.
On the other hand the spread in the occupation number is
$\Delta N=r$. The uncertainty relation is satisfied for all 
$r$, though for large $r$ it appraoches an equality.

\section{Oscillator as a planar rotor with $Z_2$ gauge invariance}
\label{z2}
In this section we point out that our adhoc procedure of extending
the Hilbert space is related to gauge invariance idea.
We use it to construct states analogous to the sine and cosine 
states of SG \cite{sg}.

We define a Large Hilbert space $\cal H$ as that of the 
planar rotor.  We define a gauge transformation, 
\begin{eqnarray}
P|n>=|-n>,n=0,\pm 1,\pm 2,\ldots
\label{29}
\end{eqnarray}
This is  the parity transformation. Our previous
technique of using only the positive Fourier modes
is releted to this formalism through a gauge fixing as 
follows.We choose a representative of each gauge equivalent
set of states to describe the physical states.Thus we may
choose $ |n>$ with $n \geq 0$ only. 

In this section we handle the gauge invariance differently.
We demand the physical states to be those that 
are invariant under this transformation.Thus the basis for the
physical states is $(|n>+|-n>)/\sqrt 2, n=1,2,3,\ldots$ in
addition to $|0>$ which is already gauge invariant.This labels the
basis for $\cal H/P$.We see that there is an one-to-one
correspondence with the basis of the oscillator,
$|0>=|0 \gg, (|n>+|-n>)/\sqrt2=|n \gg,n=1,2,3,\ldots$.

In the Large Hilbert space we have the phase eigenstate 
$|\theta>$ as in the planar rotor, Eqn.\ref{1}. However this is 
not a gauge invariant state and therefore not physical.
From Eqn.\ref{1} we see that under the gauge transformation, 
$P|\theta>=|-\theta>$.  The transformation of the wave 
function is $P \Psi(\theta)=\Psi(-\theta)$.Therefore even
periodic functions of $\theta$ are physical states.
Thus a basis for
the physical states consists of only of the even functions 
$\sqrt 2 \; cos(n\theta), n=1,2,\ldots$ and $1$ for $n=0$.

As with a particle in a box, sec. \ref{box}, the operator 
$\hat L=-i \hbar d/d\theta$ is not an operator on the physical 
states as it takes the cosine into a sine function. Since 
$P\hat L P^{-1}=-\hat L$, $\hat L$ is not gauge 
invariant.  On the other hand $\hat L^2$ is a gauge 
invariant operator and therefore well defined on the physical 
Hilbert space.Since
$\hat L^2|\pm n>=n^2 |\pm n>$, we get $\hat L^2|n\gg=
n^2|n \gg$.
On the wave functions $\Psi(\theta)$ this means $(-d^2/d\theta^2)  
cos(n \theta)=n^2 cos(n \theta), n=1,2,3,\ldots$.
Therefore we may define uniquely the square root operator,
$\hat N=\sqrt{L^2}$ which has the action
$\hat N \ cos (n \theta)=n \; cos(n \theta), \; n=1,2,3,\ldots$.
It is a self adjoint operator. This is the number operator in 
terms of which the oscillator Hamiltonian is simply 
$\hat H=(\hat N+1/2) \hbar \omega$.In this basis, the 
stationary states of the quantum oscillator have the 
form $\sqrt 2 \ cos(n \theta)$.  The time evolution is 
$\psi_n(t)=exp(-i(n+1/2) \omega t) \sqrt 2 \; cos(n \theta)$.

For any physical state the amplitude of the wave function 
at $|\theta>$ is the same as at $|-\theta>$.This means that it is 
sufficient to choose the range of $\theta$ to be $[0,\pi)$.
This may be stated in a different way.The combination
$|\theta \gg=(|\theta>+|-\theta>)/\sqrt2$ is gauge invariant and can 
be therefore expanded in terms of the complete basis
of {\sl physical} states $|n \gg$,

\begin{eqnarray}
|\theta \gg &=&\sum_{-\infty}^{\infty}\frac{1}{\sqrt 2}
(e^{in\theta}+e^{-in\theta})|n> \nonumber \\
&=& \sqrt 2(|0 \gg +\sum_{1}^{\infty} {\sqrt 2} cos(n\theta)|n
\gg)
\label{30}
\end{eqnarray}
\noindent

The state $|\theta \gg$ is not normalizable, but so is the case 
of the planar rotor where it is well understood.All operations
with this state have well defined meaning in the Large Hilbert
space.  For example,
\begin{eqnarray}
\ll \theta| \theta' \gg&=&(<\theta|+<-\theta)|
(|\theta>+|-\theta>)/2  \nonumber \\
&=&\delta^P(\theta-\theta') +\delta^P(\theta+\theta')
\label{31}
\end{eqnarray}
\noindent
As the range of $\theta$ is $[0,\pi)$ the second Dirac delta
function drops out from this equation  and we get the 
conventional orthonormality condition. 

\subsection{Alternate representation of the physical
states related to the $|\theta>$ basis}
\label{half}
The gauge invariant states $|\theta \gg$ that we constructed
from $|\theta>$ were the easiest but not the only ones.
We construct another appealing  representation.
We defined the action of the non-trivial element $P$ of the 
$Z_2$ group on the rotor states by $P|n>=|-n>$. We could 
consider other choices: $P|n>=-|-n>$.  One disadvatage with this 
is that $P|0>=-|0>$ so that it is impossible to obtain
gauge invariant states involving $|0>$.One way out is to
consider a modified planar rotor corresponding to an electric
charge confined to a circle through which is treading a half
unit of magnetic flux \cite{schu}. Now the planar angular 
momentum is $n=\pm 1/2, \pm 3/2, \ldots$.We may now choose 
the group action $P |n> = - |-n>, n=\pm 1/2, \pm 3/2, \ldots$.
Now the basis for gauge invariant states is provided by
$|n-1/2 \succ=  (|n>-|-n>)/\sqrt 2, n= 1/2, 3/2, \ldots$.
In the $|\theta>$ basis
we get the gauge invariant orthonormal set,
\begin{eqnarray}
|\theta\succ  &=& \frac{-i}{\sqrt 2}(|\theta> - 
|-\theta>) \nonumber \\
 & = & \sum_{0}^{\infty} \sqrt {2} sin((n+\frac{1}{2}) \theta)
|n \succ
\label{32}
\end{eqnarray}
\noindent
We have,
\begin{eqnarray}
\prec \theta| \theta' \succ &=&
(<\theta|-<-\theta)|(|\theta>-|-\theta>)/2  \nonumber \\
&=&\delta^P(\theta-\theta') -\delta^P(\theta+\theta')
\label{33}
\end{eqnarray}
\noindent
Again it is sufficient to restrict $\theta \ \epsilon 
\ [0, \pi)$ and we get the orthonormality.
This provides a representation of the stationary states 
of the oscillator in terms of half integer sine
functions.

\subsection{Unitary equivalence of Hilbert spaces}
\label{sg}
The cosine states of SG \cite{sg} give stationary states 
of the oscillator as those of a particle in a box. On the
other hand Eqn. \ref{30} give these wave functions as 
cosines which donot vanish at the boundary.They correspond to
stationary states of a particle in a box with 
reflecting walls which produce antinodes instead of nodes 
at the walls. Thus the same system,  viz. the oscillator, is 
being described by such disparate systems. This is yet another
instance of the unitary equivalence of all Hilbert spaces.Further
examples are the following. 
In case of our other realization of the $Z_2$ 
gauge invariance, Eqn. \ref{32}, we are realizing the
stationary states using sine functions with half integer
Fourier modes.  The sine states of SG has alternately 
sine and cosine functions for even and odd Fourier modes
respectively.

The cosine and sine states of SG\cite{sg} are the complete set
of eigenstates of hermitian cosine and sine operators. This
assures that their eigenstates form a complete orthonormal
set. In the same way, our states constructed from gauge
invariance considerations can also be realized as complete
set of eigenstates of certain Hermitian operators.These
are the operators  for which the corresponding wave 
functions are the probability amplitudes. Thus the
oscillator is being probed using different observables in
each case.In each case the eigenvalues form a continuum
bounded set. The matrix elements $\subset \theta|n>$
provide a unitary transformation ( in the generalized 
sense) from the denumerable set $|n>=0,1,2, \ldots$ to
a continuum set $|\theta) \subset [0, \pi)$ in each case.

\section{Summary}
\label{disc}
There are distinct problems to be first addressed in 
defining the phase of a quantum oscillator.We used the
planar rotor and a particle in a box to help us in resolving these
issues.These systems are relevant to the oscillator in a deeper
way.We used gauge invariance to relate it to the planar rotor
and the cosine states \cite{sg} to relate it to a particle in
a box. We defined  a new measure of phase uncertainty, Eqn.  
\ref{7}. This is closest to the r.m.s. definition without having
problems with the periodicity of the wave functions.It also
gives an uncertainty relation Eqn. \ref{12} that is of the 
conventional type with crucial differences that avoids 
contradictions.

We argued that the crucial feature required of a
conjugate variable as regards its role in uncertainty
principle is that inner product of the eigenstates of
the conjugate pair must be pure phase upto an overall
normalization. It may be necessary to extend the Hilbert
space to accomodate such an operator. We pointed out that
this is a standard procedure adopted in gauge theories.We
used this connection to justify London's phase
wave functions without running into contradictions.

We considered some examples to illustrate the way our
uncertainty relation works.
We showed that the so called coherent phase states can
be used to obtain arbitrarily narrow phase uncertainty.

\acknowledgements

I thank Professor K.H.Mariwalla and Professor R. Simon for 
useful discussions and pointing out references.

\appendix
\section{Some identities} 
\label{a}
Many of the series used here, e.g. Eqns. \ref{2},\ref{21} are not 
uniformly convergent. The ratios of successive coefficients 
are of unit modulus. Such identities can be justified 
and interpreted in the sense of generalized
functions \cite{gs}. Heuristic way of obtaining such identities 
is by regularizing the sum
and then taking the limit as described below.

The periodic delta-function can be obtained as a 
limit of the Poisson kernel \cite{j,v}  $\delta_{\epsilon}(\theta)$
in the limit $\epsilon \rightarrow 0$.
\begin{eqnarray}
\delta_\epsilon^P(\theta) 
&=&\sum_{-\infty}^{+\infty} exp(-n(\epsilon+i\theta)) \nonumber \\
&=&\frac{1-e^{-2\epsilon}}{1+
e^{-2\epsilon}-2e^{-\epsilon}cos\theta}  \nonumber \\
&&\rightarrow 0, as \ \epsilon \rightarrow 0, if 
\ \theta \neq 0(mod 2\pi), \nonumber \\
&&\rightarrow \infty, as \ \epsilon \rightarrow 0, 
if \ \theta = 0(mod 2\pi)
\label{35}
\end{eqnarray}
as required for the $\delta$ function.In particular,
$\int d\theta/(2\pi) \delta_\epsilon^P(\theta)=1$
for all $\epsilon$ including the limit $\epsilon=0$.
Thus we get $\sum_0^{\infty} 2 cos \ n\theta=
\delta^P(\theta)-1$.

We also need the following sums.
\begin{eqnarray}
\sum_{1}^{\infty} sin \ n\theta  \nonumber \\
&=& lim_{\epsilon \rightarrow 0} 
\frac{1}{2i}(\frac{1}{1-e^{-\epsilon+i\theta}} 
-\frac{1}{1-e^{-\epsilon-i\theta}})  \nonumber \\
&&\rightarrow (1/2)  cot \theta/2, as \ \epsilon \rightarrow 0, if 
\theta \ \neq 0(mod 2\pi),  \nonumber \\
&&\rightarrow 0, as \  \epsilon \rightarrow 0, 
if \ \theta = 0(mod 2\pi)
\label{36}
\end{eqnarray}
Thus there is a discontinuity at $\theta=0(mod 2\pi)$,
but we may simply take
$\sum_{1}^{\infty} 2 \ sin \ n\theta = cot(\theta/2)$.

\section{Phase operator} 
\label{b}
In sec. \ref{un} we obtained the uncertainty relation 
without explicitly introducing the operator $\hat \theta$ for the 
angular position or its commutation relation with the planar 
angular momentum.The relation can also be obtained by using these 
operators. Definition of $\hat \theta$ presents some novelty 
\cite{jl,sg,bp}. When acting on the wave function $\Psi(\theta)$ , 
$\hat \theta$ has the effect of multiplying it by a 
periodic repetition of $\theta$ in the interval 
$[-\pi, \pi)$.Therefore the action of $\hat \theta$ 
on a continuous periodic function gives a wave function that
is discontinuous at the boundary of this interval. In contrast 
the angular momentum operator $\hat L=-i \hbar d/d\theta$  has no 
such complications. This has the immediate consequence that the 
commutation relation has an extra term on the r.h.s.
\begin{eqnarray}
[\hat L,\hat \theta]=-i\hbar
(\bf{1}-\delta^P(\hat \theta-\pi))
\label{37}
\end{eqnarray}
\noindent
where the operator $\delta^P$ is obtained by using the operator 
$\hat \theta$ in place of $\theta$ in Eqn. \ref{2}.  It is this 
extra term that avoids a contradiction when we take diagonal 
matrix elements in the angular momentum basis.We get zero on both 
sides.

Our definition of the $\hat \theta$ operator selects the
$\theta =0$ point as special because our saw tooth wave
function is antisymmetric about it.We could equally well 
choose the operator $\hat \theta(\alpha)=\hat \theta+
\pi-\alpha$ which corresponds to discontinuity at 
$\theta=\alpha(mod \ 2 \pi)$.Now the commutation relation is
\begin{eqnarray}
[\hat L,\hat \theta(\alpha)]=-i\hbar
(\bf{1}-\delta^P(\hat \theta-\alpha))
\label{38}
\end{eqnarray}
\noindent
Note that $<\Psi|\delta^P(\hat \theta-\alpha)|\Psi>=
|\Psi(\alpha)|^2$.This leads to the inequality Eqn. \ref{11}.

\end{document}